\documentclass[showpacs,twocolumn,prl]{revtex4}
\usepackage{epsfig}
\usepackage{amsmath}
\usepackage{graphicx}
\usepackage{amssymb} 
\usepackage{graphics}
\usepackage{fancyhdr,graphicx}

\begin{document}

\title{Nanoscale Field Effect Transistor for Biomolecular Signal Amplification }

\author{Yu Chen$^1$, Xihua Wang$^1$, M. K. Hong$^1$, Carol Rosenberg$^2$, Shyamsunder Erramilli$^1$ \&Pritiraj Mohanty$^1$}

\affiliation{
$^1$Department of Physics, Boston University, 590 Commonwealth Avenue, Boston, MA 02215\\
$^2$Department of Medicine, Boston University School of Medicine, Boston MA 02118}


\begin{abstract}
We report amplification of biomolecular recognition signal in lithographically defined silicon nanochannel devices. The devices are configured as field effect transistors (FET) in the reversed source-drain bias region. The measurement of the differential conductance of the nanowire channels in the FET allows sensitive detection of changes in the surface potential due to biomolecular binding. Narrower silicon channels demonstrate higher sensitivity to binding due to increased surface-to-volume ratio. The operation of the device in the negative source-drain region demonstrates signal amplification. The equivalence between protein binding and change in the surface potential is described.

\end{abstract}

\maketitle

\thispagestyle{fancy}
\renewcommand{\headrulewidth}{0pt}
\fancyhead[r]{Appl. Phys. Lett. 91, 243511 (2007)}    

Ultrasensitive recognition of molecules is important in basic science, as well as in pharmacology and medicine for the analysis of biomarkers \cite{Ferr05,Grai03}. Nanotechnology has made it possible to enhance detection sensitivity by using nanowires\cite{Lieb01,Moha06}, nanotubes \cite{Zett00},nanocrystals\cite{Ping05},nanocantilevers\cite{Drav06,Moha062},and quantum dots\cite{Webb03}. The signal arising from biomolecular binding on the surface can alter measurable physical properties of the nanoscale device such as electrical conductance with increased sensitivity due to the large surface-to-volume ratio.

A silicon nanowire can be used as the source-drain channel of a field effect transistor (FET). In conventional FETs, lithographic methods are used to fabricate gates at the bottom, the top, or the side. In addition, the nanochannel surface can be functionalized with specific receptor or antibody. In a
fluid, the ligand (or antigen) can bind to the receptor, which results in a change in the surface charge profile and the surface potential. Essentially, this binding behaves as a field effect. The conductance and the I-V characteristics of the nanowire can therefore be used to characterize biomolecular binding-for instance, to determine concentration and bindin dissociation constant. In this letter, we show that characteristics
of differential conductance dI/dV can be used for even higher sensitivity in the field effect due to biomolecular binding. In particular, dI /dV characteristics allow measurement at low bias, essential for avoiding electrolysis.

In a series of experiments, Lieber and co-workers have demonstrated the operation of nanowire FET sensors fabricated with a bottom-up method, using chemically grown silicon nanowires. These nanowire field-effect sensors show significant advantages of real-time, label-free and highly sensitive detection of a wide range of analytes, including proteins, nucleic acids, small molecules, and viruses in single-element or multiplexed formats\cite{Lieb05}. These bottom-up approaches, however, have limitations arising from the lack of control in materials and fabrication engineering. In contrast, top-down methods of fabrication by e-beam lithography or submicron optical lithography enable scalable manufacturing.

Here, we demonstrate that the differential conductance dI/dV characteristics of silicon nanowire field effect transistors with width below 200 nm show strong dependence on gate voltage. We use the peak position of the dI/dV curve to characterize antibiotin detection with sub-ng/ml sensitivityby
functionalizing the silicon nanowire surface with biotin. From this measurement, we can extract the dissociation constant of biotin-antibiotin binding to be $5.2\times10^{-10}$ M, in good agreement with other measurements.

By functionalizing the silicon nanowire surface with biotin, we show that the device can be used to characterize protein binding. Furthermore, we show equivalence between the field effect created by the biomolecular binding and the reference gate voltage. This equivalence allows the characterization of ligand-receptor binding by determining the calibrated gate voltage dependence rather than measuring the more cumbersome concentration dependence.  

Our devices are fabricated by standard e-beam lithography and surface nanomachining. The starting (100) SOI wafer has a device layer thickness of 100 nm and oxide layer thickness of 380 nm, on a 600 $\mu$m boron-doped substrate. The device-layer volume resistivity ranges from 10-20 $\Omega\cdot$ cm. After patterning the nanowires and the electrode in separate steps, the structure is etched out with an anisotropic reactive-ion etch. This process exposes the three surfaces of the silicon nanowire along the longitudinal direction, resulting in structure with three-dimensional relief and a large surface-to-volume ratio. Finally, a layer of $Al_2O_3$ (5-20 nm thick) is grown by atomic layer deposition. A plasti flow cell is used to bathe the device in a small volume of fluid 20-30 (L) avoiding the mixing problem of microfluidi channel\cite{Rhee07}.

\begin{figure} [t]
	\includegraphics[scale=0.5]{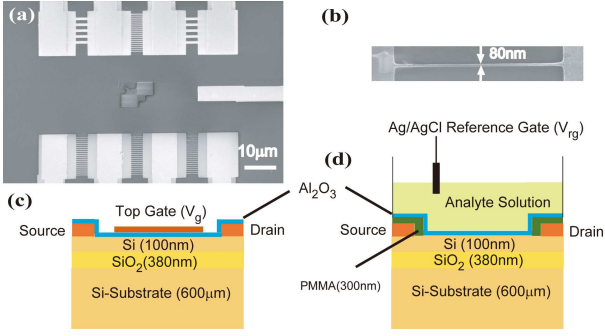}
	\caption{ Device configuration. (a) (b) Scanning electron micrographs of multiple sensor array and single wire. (c) Schematic of device cross-section view with top gate geometry. (d) Schematic of cross-sectional view of the device, showing reference gate geometry.  }
	
	\label{fig1}
\end{figure}

The device pictures and measurement configurations are shown in Fig.~\ref{fig1}. The use of parallel nanowires improves the average signal while maintaining a high surface-to-volume ratio. A fabricated top gate is used to characterize currentvoltage relation of the device in Fig.~\ref{fig1} (c). Small changes in the conductance of the device are best measured by considering the differential conductance $g(V_{ds},V_g)=\left(\frac{\partial I_{ds}}{\partial V{ds}} \right)$ with the measurements made at constant top gate voltage $V_g$. In biological sensing measurements (Fig.~\ref{fig1} (d)), an Ag/AgCl reference gate is used. The conductance is a function of the bias voltage $V_{ds}$, the gate voltage $V_g$, or equivalently the surface potential, which can be changed either by changing Vrg or the concentration of charged biomolecules\cite{Berg03}.

\pagestyle{fancy}
\fancyhead[c]{}
\rhead{Y. Chen et al, Appl. Phys. Lett. 91, 243511 (2007)}    

\begin{figure} [h]
	\includegraphics[scale=0.4]{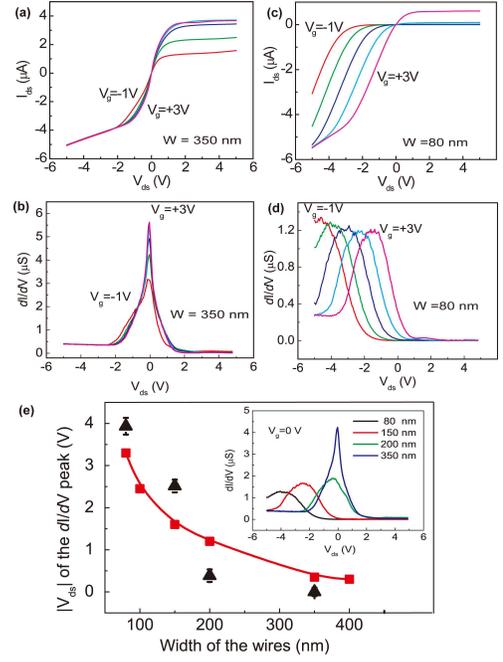}
	
	\caption{Width dependence of the I-V characteristic curve of the devices. (a) (b) show Ids and dI/dV versus source drain voltage $V_{ds}$ at different top gate voltage $V_g$ (-1 V to 3 V with 1 V step increase) for 350 nm wires. (c) (d) show $I_{ds}$ and dI/dV versus $V_{ds}$ at different $V_g$ (-1 V to  3 V with 1 V step increase) for 80 nm wires. (e) shows $\left|V_{peak}\right|$ at $V_g=0$ V for devices with different width. Red points is the simulation results used the model in reference \cite{Mohaun}. The inset is differential conductance as the function of source drain voltage at $V_g = 0$ V for different wire widths. 	}
	
	\label{fig2}
\end{figure}

Fig.~\ref{fig2} (a) is the characteristic of the device with a wire width of 350 nm. Fig~\ref{fig2}(c) shows the characteristic I-V curve for a 80 nm device. In Fig.~\ref{fig2}(b) and Fig.~\ref{fig2}(d), each curve corresponds to differential conductance measurements at fixed values of $V_g$, ranging from -1 to +3 V. At large negative $V_{ds}$, all the curves start with a common nonzero value of $g$. Such a universal value suggests that conductance is dominated by the Schottky barrier at the junction\cite{Sze81}. Fig.~\ref{fig2}(b) shows differential conductance for the 350 nm wire as a function of Vds, which shows a peak in differential conductance near 0 V. As Vds is increased above a threshold voltage $V_{th}$, the slopes of both I-V characteristic and differential conductance increase. For the 350 nm wire, the differential conductance peak positionin Fig.~\ref{fig2}(b) is insensitive to $V_g$. In contrast, the peak position in Fig.~\ref{fig2}(d) is strongly dependent on $V_g$. The sensitivity of the conductance peak of the device with narrower width comes from the threshold voltage change. In our model, $V_{th}$ is linearly dependent on the surface-to-volume ratio

\begin{equation}
	\frac{A_{sur}}{V_{vol}}=\frac{(w+2H)}{wHL}=\frac{w+2H}{wH}.
\end{equation}
Here, w is the width of the wire, H is the thickness of the device layer, and L is the length of the wire. The peak position of the differential conductance at zero top gate voltage is plotted as a function of the nanowire width in Fig.~\ref{fig2}(e). Following the methods given in Ref.\cite{Mohaun}, the conductance peak of the dI/dV versus Vds can be calculated in a semiconductor model. With increased surface-to-volume ratio, and hence a higher threshold voltage, the conductance channel is closed at zero bias. A higher gate voltage, i.e., a larger negative bias voltage, is then needed to open the narrow wire channel. Therefore, the conductance peak shifts to negative drain voltage (Fig.~\ref{fig2}(e), red points).

In the experiment, the reference gate is immersed in the solution, and a back gate voltage is applied to the substrate. In an array of sensors, the solution reference gate voltage $V_{rg}$ is common to all the nanowires. However, the source-drain voltage $V_{ds}$ for each nanowire in the nanosensor array can be controlled individually. In this more flexible configuration, the high voltage required for the back gate can be avoided\cite{Rhee07}. At the same time, after channel is opened gradually by $V_{ds}$, the sensitivity of the device can be amplified. This naturally implies a region of operation at negative bias for high sensitivity measurements.

\begin{figure} [h]
	\includegraphics[scale=0.5]{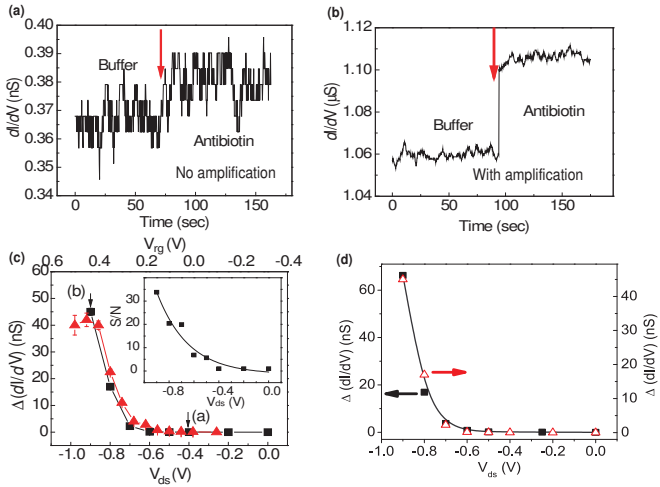}
	\caption{ (a) (b) Differential conductance versus time. The phosphate buffer is first injected, followed by a 100 ng/mL of anti-biotin in same buffer, with data acquisition being stopped for 30 s following the injection. The bias is set at $V_{ds}=-0.4$ V and $V_{ds}= -0.9$ V, respectively.  (c) Change in the differential conductance of the device introduced by antibiotin at different Vds (bottom axis, with fixed $V_{rg}=0.3$ V, black solid squares). Red solid triangles are the data at different Vrg (top axis, with $V_{ds} = 0$ V. Inset is the signal noise ratio of the device at different $V_{ds}$. (d) Comparison of conductance change  introduced by 5 mV of reference gate voltage change(black dots, left axis) and 100 ng/mL of antibiotin solution (red triangles, right axis).}
	\label{fig3}
\end{figure}

The signal amplification effect in reverse bias is demonstrated in protein (antibiotin) measurements. Silicon nanowires, functionalized with biotinylated bovine serum albumin, are used to detect antibiotin in 1 mM NaCl and 1 mM phosphate buffer solution. At the concentration of salt used
in solution, the Debye screening length $\lambda_d$ at room temperature is about 9.6 nm. $\lambda_d$ is sufficiently large that the surface potential is sensitive to protein binding, but short enough to screen out biomolecules in solution.11,15 All solutions used in the measurements are dialyzed to maintain constant ionic strength. The differential conductance change $\Delta g$ due to 100 ng/ml antibiotin in the solution is $0.02\pm0.01$ nS at $V_{ds} =-0.4$ V (see Fig.~\ref{fig3}(a)), while $\Delta g$ is $45\pm0.1 $ nS at $V_{ds} =-0.9$ V (see Fig.~\ref{fig3}(b)). The signal-to-noise ratio increasesfrom 2 to 34 (see Fig.~\ref{fig3}(c) inset). The signal due to 100 ng/ml antibiotin injection is plotted as a function of Vds and $V_{rg}$ (see Fig.~\ref{fig3}(c)). This plot clearly shows the effect ofreverse bias amplification.

The change $\Delta g$ above, due to concentration change at fixed reference gate voltage, can be compared to the change caused by varying the reference gate voltage while keeping the concentration fixed. An equivalence between the surface potential change and concentration change is then established.
$\Delta g$ introduced by 100 ng/ml ((660 pM) of antibiotin is equivalent to a gate voltage change of $7.2\pm0.3$ mV(see Fig.~\ref{fig3}(d))

The fundamental advantage of our label-free device architecture is the combination of high detection sensitivity and standard semiconductor-based fabrication techniques, suitable for scalable manufacturing. We operate the device in the reverse bias region without applying high voltage to the
back gate\cite{Rhee07} or to the reference gate voltage in the solution. Our device configuration allows device-level signal amplification and the required degree of control to enable largescale parallel architecture for detection of multiple target molecules. This combination is important for any clinical application, including disease screening, diagnosis, monitoring,and prognosis, all of which require fast analysis time, small sample volume, and low cost.

In conclusion, we fabricate silicon nanowire channel FET sensors by a top-down e-beam lithography method. Differential conductance characteristics of the devices show dependence on the source-drain channel width of the nanoscale sensor. We take advantage of the strong dependence in the reverse-bias region to demonstrate amplification of biomolecular binding signal by modulating the top gate voltage. The operation of the device at low reference gate voltage prevents electrolysis. By comparing the signal generated by protein binding with the change in the reference gate voltage, we establish an equivalence between the effect arising due to the concentration-dependent surface charge density and the tunable surface potential in the silicon nanowires.

The authors acknowledge support from the Department of Defense and the National Science Foundation and this work is performed in part at the Photonics Center.


\end{document}